\documentclass[preprint]{emulateapj} 

\shorttitle{Cold Classical KBOs are Primordial} 
\shortauthors{Batygin Brown Fraser} 

\begin{document}
 
\title{Retention of a Primordial Cold Classical Kuiper Belt in an Instability-Driven Model of Solar System Formation}  
\author{Konstantin Batygin, Michael E. Brown \& Wesley C. Fraser} 

\affil{Division of Geological and Planetary Sciences, California Institute of Technology, Pasadena, CA 91125}

\email{kbatygin@gps.caltech.edu}

\begin{abstract}
The cold classical population of the Kuiper belt exhibits a wide variety of unique physical characteristics, which collectively suggest that its dynamical coherence has been maintained through out the solar system's lifetime. Simultaneously, the retention of the cold population's relatively unexcited orbital state has remained a mystery, especially in the context of a solar system formation model, that is driven by a transient period of instability, where Neptune is temporarily eccentric. Here, we show that the cold belt can survive the instability, and its dynamical structure can be reproduced. We develop a simple analytical model for secular excitation of cold KBOs and show that comparatively fast apsidal precession and nodal recession of Neptune, during the eccentric phase, are essential for preservation of an unexcited state in the cold classical region. Subsequently, we confirm our results with self-consistent N-body simulations. We further show that contamination of the hot classical and scattered populations by objects of similar nature to that of cold classicals has been instrumental in shaping the vast physical diversity inherent to the Kuiper belt.

\end{abstract}

\section{Introduction}
The quest to understand the origins of the solar system dates back centuries. The last two decades, however, have seen a renewed interested in the problem, as the discovery of the Kuiper belt \citep{1993Natur.362..730J} has provided important new clues to the physical processes that took place during the early stages of our solar system's evolution. The continued acquisition of new information gave rise to a multitude of new formation models (see \cite{2008ssbn.book..275M}  for a comprehensive review). Among the newly proposed scenarios, an instability model, termed the ``Nice" model \citep{2005Natur.435..459T, 2005Natur.435..466G, 2005Natur.435..462M}, has been particularly successful in reproducing the observed properties of planetary orbits and the Kuiper belt \citep{2008Icar..196..258L}.

Within the context of the narrative told by the Nice model, planets start out in a multi-resonant configuration \citep{2007AJ....134.1790M, 2010ApJ...716.1323B}, and, driven by planetesimal scattering, begin migrating divergently \citep{1984Icar...58..109F}. Eventually, the planets encounter a low-order mean motion resonance, which results in a transient period of instability. During this period, the ice giants scatter outwards and settle roughly onto their current semi-major axes but with high eccentricities \citep{2005Natur.435..459T, 2008ApJ...675.1538T}. Neptune's excited eccentricity gives rise to a chaotic sea between its exterior 3:2 and 2:1 mean motion resonances (MMRs), allowing planetesimals to random-walk into the ``classical`` region \citep{2008Icar..196..258L}. Subsequently, as the planets circularize due to dynamical friction \citep{1988Icar...74..542S}, the scattered and resonant populations of the Kuiper belt are sculpted.

An outstanding problem within the Nice model lies in the formation of the cold classical population of the Kuiper belt, which is the central theme of this study. The cold population is distinctive from the rest of the Kuiper belt in a number of ways. First and foremost, as its name suggests, the orbital distribution is dynamically unexcited. When Neptune scatters planetesimals, it tends to pump up their inclinations to tens of degrees. Yet the cold population resides on nearly co-planar orbits, with inclinations not exceeding $\sim 5 \deg$ \citep{2001Icar..151..190B, 2008ssbn.book...43G}. The eccentricities of the cold population, on average, also tend to be diminished in comparison with the hot population, but the division there is not as apparent. Figure 1 shows the eccentricities of the current aggregate of observed Kuiper Belt objects (KBOs) between $30$ and $60$AU. Cold classical objects, whose inclinations are below $5 \deg$ are plotted as black dots, while all other objects with inclinations above $5 \deg$ are plotted as blue dots. Note that cold population's eccentricity distribution is not monotonic in semi-major axes. Between $42$AU and $45$AU, planar KBOs have roughly isentropic eccentricities. However, low-eccentricity objects progressively disappear beyond $45$AU. We refer to this feature of the Kuiper belt as the ``wedge" (see Figure 1).

A second distinction is the colors of cold classical KBOs. In general, the Kuiper belt exhibits a vast diversity of colors, from neutral gray to deep red. Within this range, cold classical KBOs readily stand out as clump of exclusively red material \citep{2002ApJ...566L.125T, 2005EM&P...97..107L}.  In a similar manner, the size distribution of the cold population differs significantly from that of the hot classical population \citep{2010Icar..210..944F}. Finally, the fraction of binaries present in the cold population is uniquely large \citep{2006AJ....131.1142S}. Moreover, it has been shown that the wide binaries of the cold population in particular, would have been disrupted by encounters with Neptune \citep{2010ApJ...722L.204P}, and thus must have never been scattered. While it is difficult to interpret each of these observational facts as conclusive evidence for a particular history, their coherence suggests that the cold classicals are a unique population whose dynamical similarity has been maintained through the dramatic evolution of the outer solar system \citep{2004come.book..175M}.

A number of formation mechanisms for the cold classical population have been suggested. Within the context of a smooth migration scenario \citep{1995AJ....110..420M, 2005ApJ...619..623M, 2005AJ....130.2392H}, a primordially cold population can in principle escape dynamical excitation. However, other drawbacks of the smooth migration scenario, such as the inability to reproduce secular architecture of the planets and difficulties in forming the hot classical belt, render it unlikely \citep{2009A&A...507.1041M, 2009A&A...507.1053B}. \cite{2008Icar..196..258L} advocated a similar emplacement history for the cold classicals as the hot classicals (i.e. via MMR overlap). The cold population that is produced in such simulations however, is not cold enough and not physically distinct from the hot population. Subsequently, \cite{2008ssbn.book..275M} showed that if a local, cold population is implemented into the orbital solution of \cite{2008Icar..196..258L}, it will have the same orbital distribution as the implanted population after the instability, so the problem remains. Thus, no coherent picture of the formation of the cold population exists.

\begin{figure}[t]
\includegraphics[width=0.5\textwidth]{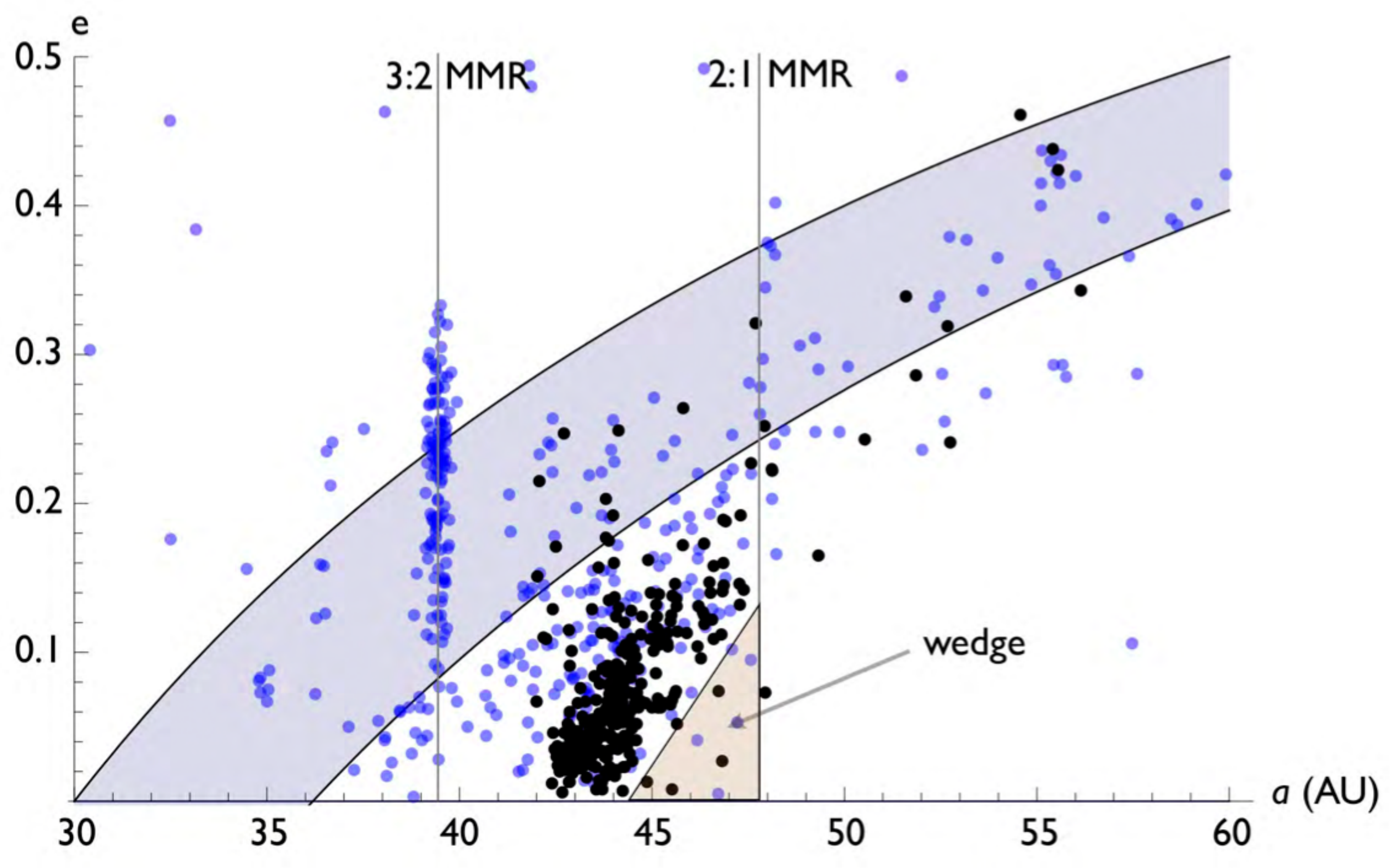}
\caption{Semi-major axis vs. eccentricity of he observed Kuiper belt. The black points denote objects with inclinations below $i < 5 \deg$ i.e. the cold classical population. The blue points represent all other objects with $i > 5 \deg$. The filled curves represent the scattered disk region and the major mean-motion resonances are shown as solid lines. The triangle, adjacent to the 2:1 MMR depicts the wedge structure, inherent to the cold classical population.} 
\end{figure}

In this work, we show that in-situ formation of the cold population is consistent with an instability model and all observed dynamical properties of the population, including the low inclinations and the wedge (shown in Figure 1) can be formed. The aim here is not to replicate the Kuiper belt and the orbits of the planets in a complex N-body simulation, but rather to identify the dynamical processes responsible for the sculpting of the region. The plan of our paper is as follows: in section 2, we construct an analytical model for secular excitation of a primordially unexcited belt, and thus derive the conditions for retention of dynamically cold orbits. Moreover, we show that the now-fossil wedge is a result of a temporary slow-down in orbital precession. In section 3, we perform self-consistent N-body simulations that confirm our analytical results. Numerical simulations show that while the cold population can remain undisrupted, similar objects immediately interior to the 3:2 MMR get scattered all over the Kuiper belt. We conclude and discuss our results in section 4. 

\section{Secular Excitation of the Cold Kuiper-Belt}

Here, we seek to develop a simplified analytical model that describes the long-term interactions between Neptune and an initially dynamically cold population of KBOs, residing between its exterior 3:2 and 2:1 MMRs, during a transient phase of high eccentricity. Prior to the instability, the planets sit in a compact configuration on near-circular orbits. As long as the orbital separation between the planets and KBOs remains large, their mutual interactions are extremely weak, so the KBOs maintain their dynamically cold orbits. Consequently this period is unimportant to the problem at hand. 

When planet-planet resonance crossing (or some other mechanism) causes the instability, the gain in semi-major axes and acquisition of high eccentricities and inclination of the planets takes place on a time scale that is considerably shorter than Neptune's apsidal precession period (i.e. less than a million years or so). As a result, it can be viewed as instantaneous within the context of a secular approximation. Thus, in an orbit-averaged sense, it is as if Neptune suddenly appears at $30$ AU with a high $e$ and $i$ and begins interacting with the KBOs. Since we seek to show that after the transient phase of high eccentricity, the KBOs can end up on dynamically cold orbits, we must restrict Neptune from penetrating the region beyond $40$ AU. This places the maximum eccentricity, attainable by Neptune, below $e_{max} < (4/3 -1) =1/3$. This is however a weak constraint, since an eccentric, inclined Neptune can still cause large modulations in the eccentricities and inclinations of the KBOs on a secular time-scale \citep{1999ssd..book.....M}. Let us now develop a mathematical model for these secular interactions.

We begin by modeling Neptune's evolution. In our model, we take the mass of the cold KBOs to be negligible, so  they have no effect on Neptune's orbit (this is not necessarily true, at all times, for other Kuiper belt populations). The lack of mass in the primordial cold belt is a requirement for our model that brings up concerns about its formation. We shall discuss this in some detail in section 4. Since we seek to retain the majority of the local population, and we know that the mass of the current cold classical population is much less than that of the Earth, this is a reasonable assumption. The other planets, as well as the massive component of the Kuiper belt, will cause apsidal and nodal precession of Neptune's orbit, which we write as $g = \left<\dot{\varpi}_N \right>$ and $f= \left<\dot{\Omega}_N\right>$ respectively. Note that we are only accounting for the average precessions. We express dynamical friction as exponential decay of $e$ \& $i$ with constant time-scales $\tau_e$ and $\tau_i$. These time-scales are different, and their numerical values in N-body simulations tend to be of order $\sim 10^7$ years \citep{2008Icar..196..258L}. We neglect the modulation of Neptune's $e$ \& $i$ by the other planets. In other words, we only retain the free elements. 

In terms of complex Poincar$\mathrm{\grave{e}}$ variables ($x = e \exp(\imath \varpi), y = i \exp(\imath \Omega)$), we can formulate the first-order Lagrange's equations for Neptune as follows:
\begin{equation}
\frac{d x_n}{d t} = \imath g x_n - \frac{x_n}{\tau_e} \ \ \ \   \frac{d y_n}{d t} = \imath f y_n - \frac{y_n}{\tau_i}. 
\end{equation}
\\
It is trivial to show that these equations admit the solutions
\begin{equation}
x_n = e_n^0 \exp[(\imath g - 1/\tau_e)t ]  \ \ \ \ y_n = i_n^0 \exp[(\imath f - 1/\tau_i)t ] ,
\end{equation}
\\
where $e_n^0$ and $i_n^0$ are the initial (maximum) eccentricity and inclination of Neptune respectively and $\imath = \sqrt{-1}$. Here, we take $e_n^0=0.25$ and $i_n^0=10 \deg$, in accord with results of numerical simulations \citep{2005Natur.435..459T, 2008Icar..196..258L, 2010ApJ...716.1323B}. In our simple model, the secular evolution of the KBOs is dictated entirely by Neptune's evolution. In the spirit of Laplace-Lagrange secular theory, we only retain terms up to second order in eccentricity and inclination in the disturbing function of the KBOs to ensure a decoupled, analytical solution. The resulting first-order Lagrange's equations read \citep{2002ApJ...564.1024W}
\begin{equation}
\frac{d x_{kbo}}{d t} = \imath A x_{kbo} + \imath A_{n} x_n \ \ \ \ \frac{d y_{kbo}}{d t} =  \imath B y_{kbo} + \imath B_{n} y_n,
\end{equation}
where $A$, $A_n$, $B$, and $B_n$ are constants that depend only on the planetary masses and semi-major axes ratios of Neptune to KBOs (e.g. Ch.7 of \cite{1999ssd..book.....M}). Note that in the free precession terms, ($A$,$B$), the presence of other planets can also be accounted for with ease. 

\begin{figure}[t]
\includegraphics[width=0.5\textwidth]{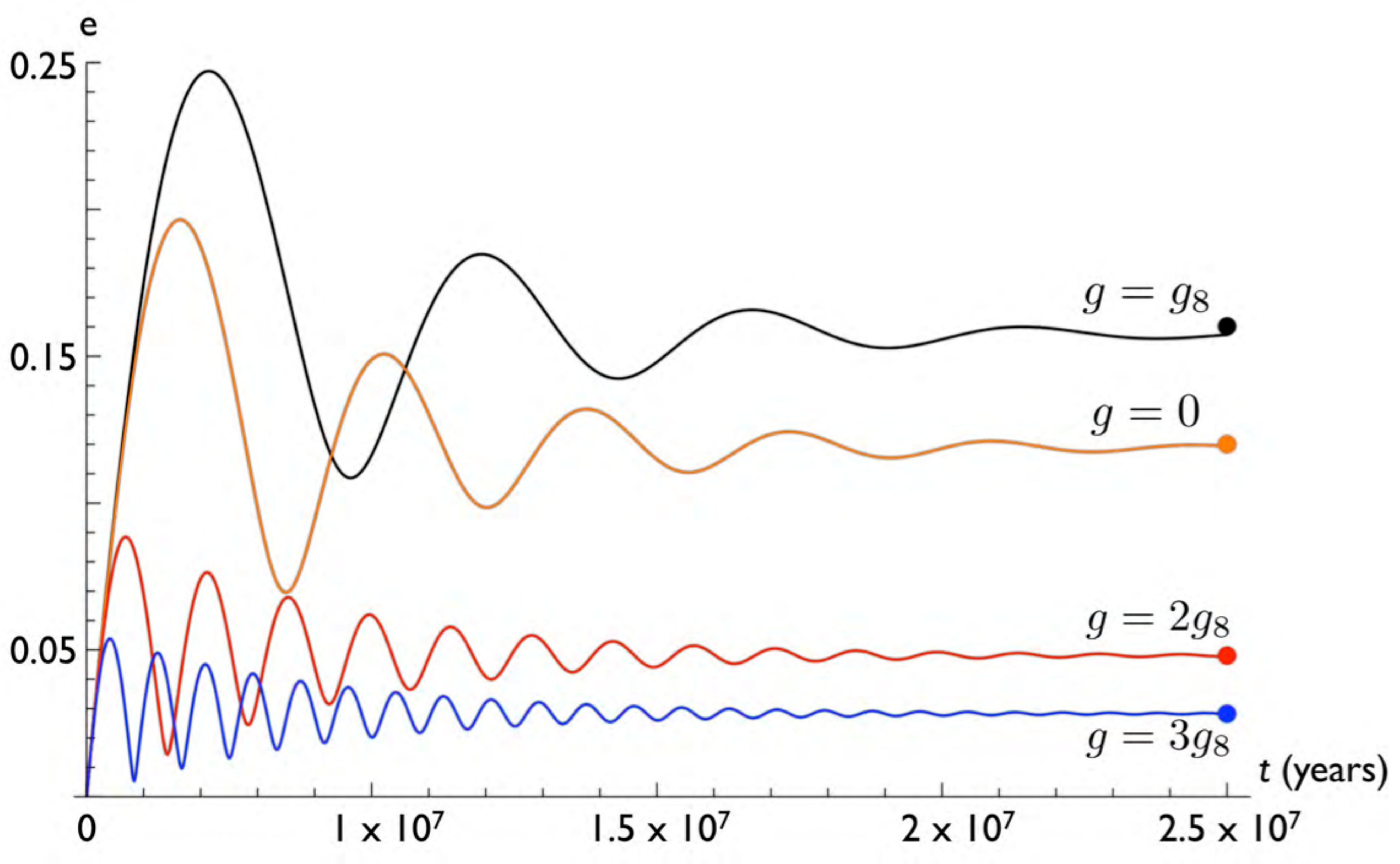}
\caption{Secular excitation of a KBO at $a=45$AU, as dictated by equation (4). In these solutions, we chose $a_N=30$AU, $e_n^0=0.25$, and $\tau_e=4$Myr. The final eccentricities (given by equation 8) are plotted as dots. Note that a low final eccentricity requires a comparatively fast precession.} 
\end{figure}

From here, let us focus only on the eccentricity evolution, since the derivation of the inclination evolution follows an identical procedure. Setting the initial orbital state vector the KBO to zero ($\left[ x, y \right] = \vec{0} $), the solution to the above equation reads
\begin{equation}
x_{kbo}=\frac{e_n^0 \tau_e A_{n} ( \exp[\imath A t] -  \exp[(\imath g - 1/\tau_e) t] ) }{A \tau_e -g \tau_e  -\imath}.
\end{equation}
The controlling parameter in this solution is Neptune's precession, $g$. Four solutions, for a KBO at $a=45$AU, with different $g$'s are presented in Figure 2. A natural unit of $g$ is the $g_8$ eigenfrequency of the Laplace-Lagrange secular solution for the solar system, which physically corresponds to Neptune's average precession rate in the current solar system (see \cite{1999ssd..book.....M}). Incidentally, the same unit can be used for the nodal recession rate in the inclination solution, since quantitatively $g_8 \approx -f_8 \approx 0.65 "/yr$. As can be seen in Figure 2, varying $g$ leads to dramatically different results. In particular, if low-eccentricities are to be retained,  $g$ must significantly exceed $g_8$. 

After a sufficient amount of time, when Neptune's eccentricity has decayed away (i.e. $t \gg \tau_e$), the second exponential in the numerator of equation (4) can be neglected. Such a solution represents a precessing KBO with a constant eccentricity. Accordingly, the time dependence of the solution only governs its angular part. Since we are solely interested in the final orbits of the KBOs, we must extract only the radial part of the solution. Let us write the $t \gg \tau_e$ solution as an exponential of an arbitrary number, $\xi$:
\begin{equation}
\exp(\xi)=\frac{e_n^0 \tau_e A_{n} \exp[\imath A t]  }{A \tau -g \tau  -\imath}.
\end{equation}
Solving for $\xi$, and complex-expanding the logarithm, we have
\begin{eqnarray}
\xi&=&\ln( \frac{e_n^0 \tau_e A_{n} }{\sqrt{1+\tau_e^2 (g - A )^2 }} ) \nonumber \\
&+& \imath \arg(-\frac{e_n^0 \tau_e A_{n}  \exp[\imath A t]  }{\imath - \tau_e (A-g)}).
\end{eqnarray}
The argument of the logarithm in the above equation is the radial part of the complex solution, which corresponds to the final eccentricity of the KBO, with an equivalent expression for the inclination:
\begin{eqnarray}
e_{kbo}^{final}=\frac{e_n^0 \tau_e A_{n} }{\sqrt{1+\tau_e^2 (g - A )^2 }} \nonumber \\ 
i_{kbo}^{final}=\frac{i_n^0 \tau_i B_{n} }{\sqrt{1+\tau_i^2 (f - B )^2 }}.
\end{eqnarray}
\\
In principle, we could have arrived at the same answer by complex-expanding the solution and taking the square root of the sum of the squares of the real and imaginary parts, although the intermediate expressions would have been considerably more messy.

The above equations can be simplified even further by considering their limiting regimes. If the decay time-scale is much longer than the beat frequency $(g-A,f-B)$, we can Taylor expand the equations to first order in ($1/\tau$) around zero. The answer then becomes independent of $\tau$. 
\begin{eqnarray}
e_{kbo}^{final} \simeq e_n^0 \frac{ A_{n} }{ g - A  } \ \ \ \ i_{kbo}^{final} \simeq i_n^0 \frac{B_{n} }{f - B }
\end{eqnarray}
This procedure is equivalent to assuming that $\tau_e^2 (g - A )^2 \gg 1$ or $\tau_i^2 (f - B )^2 \gg 1$ and throwing away the $1$ under the square root in the denominator\footnote{Alternatively, if the decay time-scales are short, we are in the non-adiabatic regime, where the solutions become $e_{kbo}^{final} \simeq e_n^0 \tau_e A_{n}$ and $i_{kbo}^{final} \simeq i_n^0 \tau_i B_{n} $}. It is clear from Figure 2, where the approximate solutions are plotted as big dots, that quantitative agreement with the ``full" solution (equation 4) is excellent, in the parameter regime of interest.

Figures 3 and 4 show the secular excitation of initially cold KBOs eccentricities and inclinations between the 3:2 and 2:1 MMRs with $g=0,g_8,2g_8$ and $3g_8$ as solid lines. These solutions suggest that if one is to retain an eccentricity below $e < 0.1$ and inclination below $i < 5 \deg$, Neptune's average orbital precession and nodal recession rates must have exceeded $\sim 3 g_8$ during the eccentric phase. The enhanced precession is primarily a consequence of Uranus. When Neptune scatters, it arrives somewhat closer to the sun than its current orbit and migrates to $\sim 30$AU by scattering KBOs (here, we have implicitly omitted this effect by stating that the coefficients $A$, $A_n$, $B$, and $B_n$ are constant). Thus, at the time of scattering, the semi-major axis ratio of Neptune to Uranus may be lower, leading to an enhanced precession. Additionally, the mass contained in the Kuiper belt may also play a role in inducing secular precession of Neptune.

The solution described above gives eccentricities that monotonically decrease with semi-major axes. However, as already discussed above, the observed cold population exhibits a somewhat different behavior, with low-eccentricity objects progressively disappearing in the vicinity of the 2:1 MMR. This dynamically unique structure (i.e. the wedge - see Figure 1) is an essential feature to any proposed formation mechanism for the cold classicals. 

\begin{figure}[t]
\includegraphics[width=0.5\textwidth]{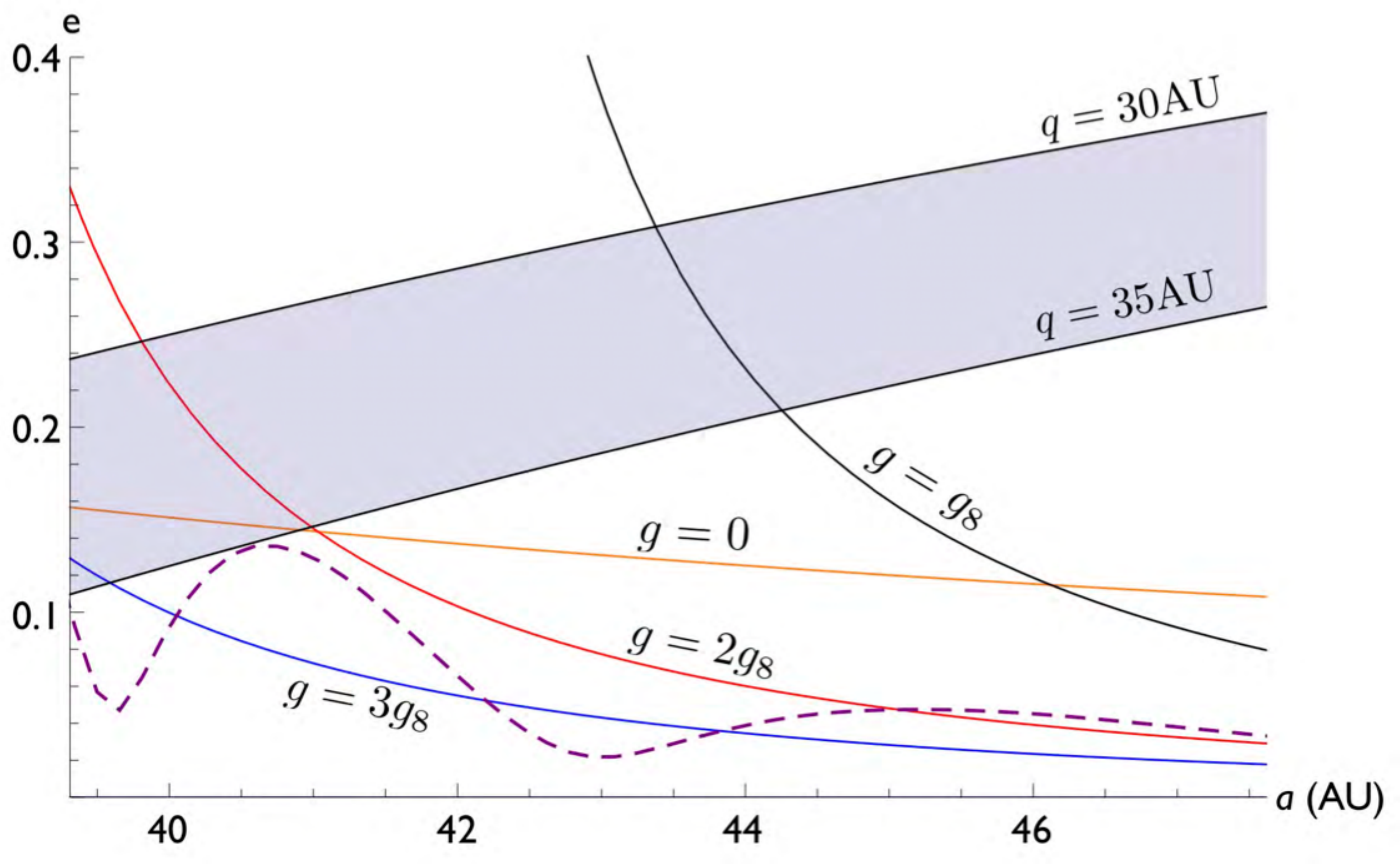}
\caption{Post-excitation (final) eccentricities in the cold region of the Kuiper belt. Solutions with $g=0,g_8,2g_8$ and $3g_8$ are presented as solid lines. Note that in order to retain nearly-circular orbits $g \gtrsim 3 g_8$ is required. The dashed line represents a solution where Neptune's precession rate is not kept constant. The shaded region corresponds to the scattered disk.} 
\end{figure}

A wedge-like structure cannot be reproduced by a sweeping 2:1 MMR in an instability-driven formation model. Unlike the smooth migration scenario, where resonant capture is possible \citep{1983CeMec..30..197H, 1995AJ....110..420M, 2011ApJ...726...53K}, when Neptune is eccentric, the chaotic motion, that arises from resonant splitting \citep{1980AJ.....85.1122W}, leads to an effective randomization of the eccentricities \citep{2006MNRAS.372L..14Q}. In other words, the KBOs that are temporarily captured do not form a coherent structure such as the wedge. An alternative scenario for formation of the wedge is one where the local population ends at $45$AU, and the wedge is a result of an extended scattered disk with $q \sim 40$AU \citep{2008ssbn.book...43G}. It is unlikely, however, that in the extended scattered disk scenario, the low inclinations of scattered objects could be preserved.

Here, we propose the formation of the wedge to be a consequence of secular perturbations. Thus, we seek to modify the above secular solution such that it yields eccentricities that are not monotonically decreasing with semi-major axes in the region of interest. As already described above, the controlling parameter in the secular solution is $g$. So far, we have kept $g$ constant. However, since Neptune scatters numerous KBOs during its circularization, and the orbits of other planets (particularly Uranus) are changing as well, one would expect Neptune's precession to vary considerably, in a chaotic manner.

It is difficult to predict the exact nature of this variation without a detailed calculation, so here we consider an extreme case as a proof of concept. Namely, we set $g = 4 g_8$ at all times, except $\tau < t < 1.1\tau$, where we set $g = 0$. Note that the precession of Neptune need not necessarily stop. We are choosing $g = 0$, rather than a diminished precession rate (such as, say $g = g_8$) merely for the sake of argument. An analytical solution is attainable in a similar fashion as above, by breaking up the integration into $3$ separate time intervals. If $g$ is not held constant through out Neptune's circularization, the final eccentricity and inclination take on a different character. Qualitatively, this can be understood as follows: when Neptune stops precessing, it starts to induce considerable oscillations in eccentricities of KBOs; however, once the precession becomes rapid again, the modulation stops and the eccentricities become frozen-in. These solutions are plotted as dashed curves in Figures 3 and 4. The details of the non-monotonic solution depend on when and for how long Neptune's precession is halted, and change further if the precession is merely slowed down, rather than stopped. Furthermore, the dashed curves in Figures 3 and 4 shift to larger semi-major axes if the free precessions of the KBOs ($A, B$) are enhanced. While it is understood that these calculations do not reproduce the cold classical population in detail, they do show that primordially unexcited objects can retain cold orbits in face of dynamical excitation, and coherent structure can be formed in the context of a purely secular solution.

\begin{figure}[t]
\includegraphics[width=0.47\textwidth]{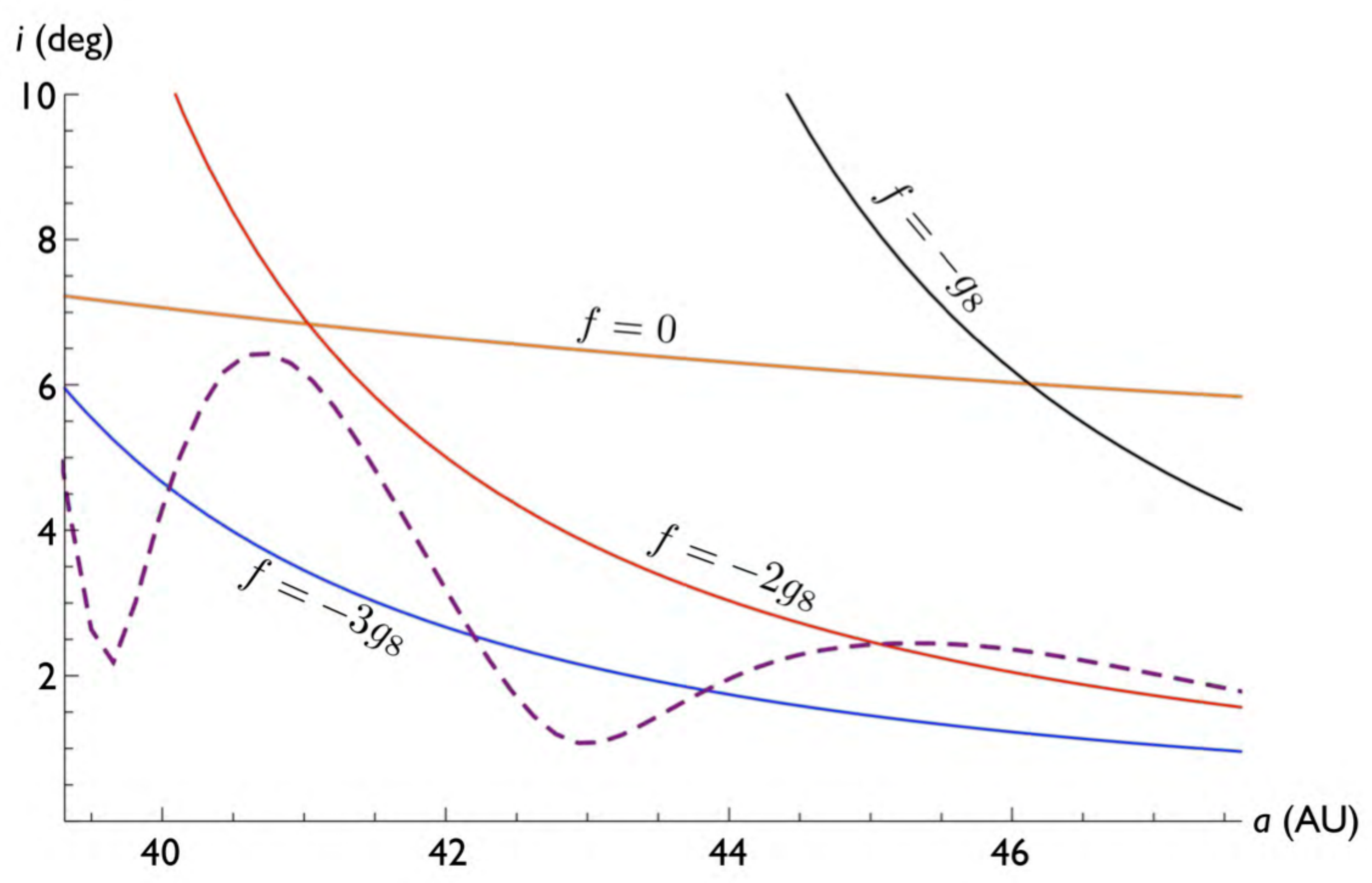}
\caption{Post-excitation (final) inclinations in the cold region of the Kuiper belt. Solutions with $f=0,g_8,2g_8$ and $3g_8$ are presented as solid lines. Note that in order to retain $i \lesssim 5\deg$ in the $42-45$AU region $f \lesssim - 3 g_8$ is required. The dashed line represents a solution where Neptune's nodal recession rate is not kept constant. Note that the quantitative character of the solution here is subtly different from the eccentricity solution (Figure 3). This is because $B$ involves a Laplace coefficient of the first kind, while $A$ involves that of the second kind.} 
\end{figure}

\section{Numerical Simulations}

Having motivated in-situ formation of the cold classical population with analytical arguments, we now turn to numerical N-body simulations for confirmation of the above results and inclusion of omitted physics (such as close-encounters, mean-motion resonances, and higher-order secular terms in the disturbing function). In this study, the integrations were performed using the \textit{mercury6} integration software package \citep{1999MNRAS.304..793C} utilizing the ``hybrid" algorithm. The disk was composed of two components. The massive planetesimal swarm, containing $3000$ particles, resided between the immediate stability boundary of the initial multi-resonant configuration and $\sim 35$AU. This was followed by a disk of another $3000$ mass-less particles that extended to $60$ AU. Thus we are assuming that a significant density gradient exists, in the vicinity of Neptune's final orbit, such that the mass in the outer disk is insufficient to drive Neptune's migration. However, the numbers of particles were chosen due to considerations of computational cost and are not intended to be representative of the relative fraction of bodies in the planetesimal disk in any way. The initial conditions were drawn from the eight multi-resonant states that were identified by \citet{2010ApJ...716.1323B} as being compatible with an instability formation model. The planetesimals were initialized on near-coplanar, near-circular orbits ($e \sim \sin i \sim 10^{-3}$). The self-gravity of the planetesimal swarm was neglected to reduce computational cost of the experiments, as 30 permutations of each initial condition were integrated. 

\begin{figure}[t]
\includegraphics[width=0.5\textwidth]{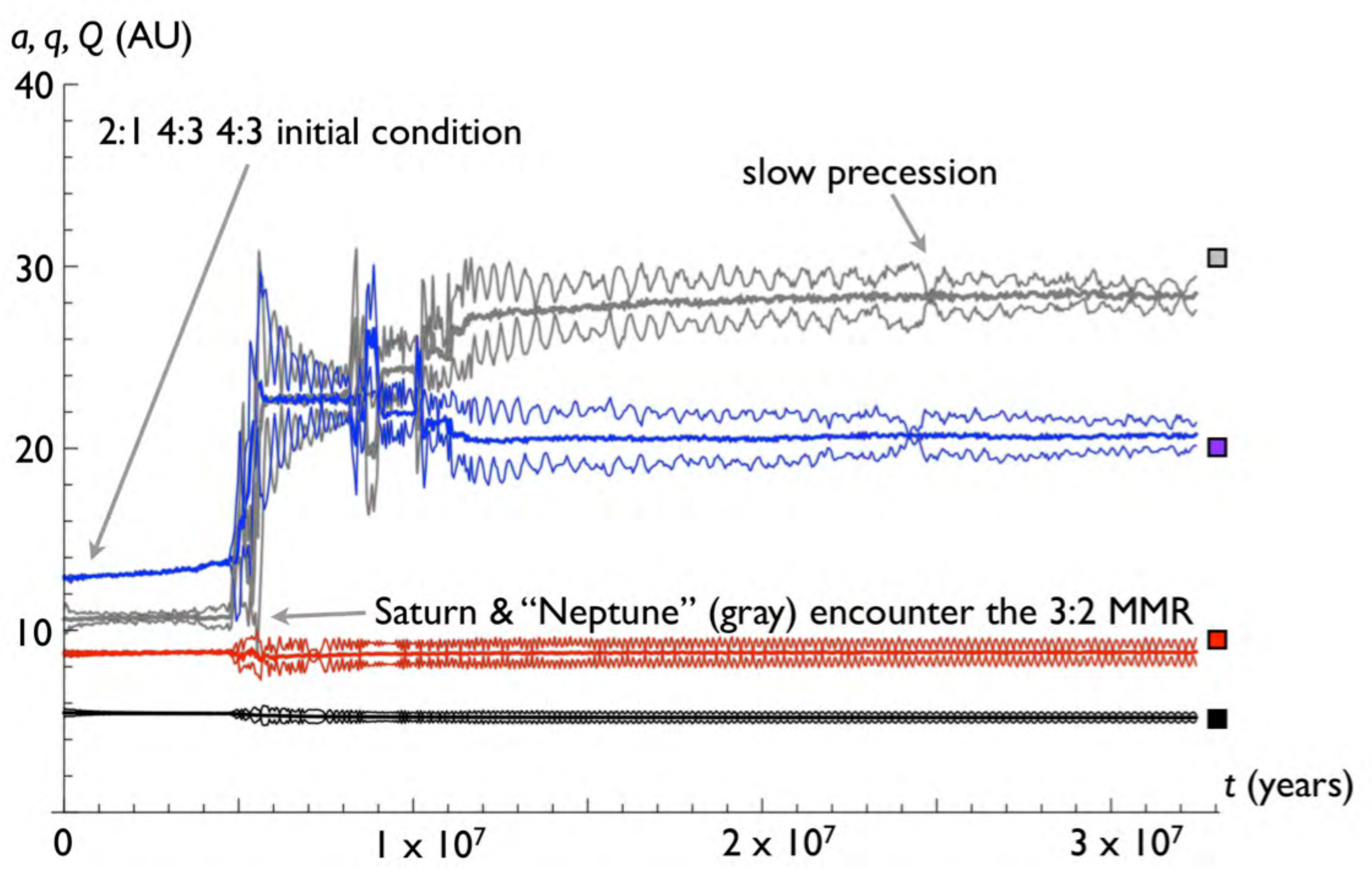}
\caption{Orbital evolution of planets. The system starts out in the (2:1 4:3 4:3) initial condition, and undergoes a brief period of instability when Neptune (gray) and Saturn (Red) encounter a mutual 3:2 MMR. At $t \approx 22$Myr, Neptune's precession rate temporarily slows down and sculpts the wedge. The boxes on the right of the plot correspond to actual semi-major axes of the giant planets. An evolved remnant planetesimal disk of this simulation is presented in Figures 7 \& 8.} 
\end{figure}

\citet{2010ApJ...716.1323B} used the presence of scattering events between an ice giant and a gas giant, followed by a transient phase of high eccentricity, as a proxy for whether successful formation of the classical Kuiper belt can occur. Further constraints on the initial conditions can be placed by considering the reproduction of the outer solar systems' secular eigenmodes. Particular difficulty has been found in ensuring that the amplitude of Jupiters $g_5$ mode is larger than that of the $g_6$ mode \citep{2009A&A...507.1041M}. Having completed all of the integrations, we checked the relative amplitudes of the $g_5$ and the $g_6$ modes in all solutions. Surprisingly, we found that despite a transient period of instability and gas giant/ice giant scattering, the (3:2 3:2 4:3)\footnote{In our notation, each pair of numbers represents an MMR in the multi-resonant initial condition. For example, (3:2 3:2 4:3) corresponds to a starting state where Jupiter \& Saturn, as well as Saturn \& Uranus are in 3:2 MMRs, while Uranus \& Neptune are in a 4:3 MMR.} and the (5:3 4:3 4:3) initial conditions did not reproduce the secular architecture of the planets, in neither this set nor in the set of integrations of \citet{2010ApJ...716.1323B}. If Jupiter and Saturn were indeed initially locked in the 3:2 MMR, as hydrodynamic simulations suggest \citep{2001MNRAS.320L..55M, 2007Icar..191..158M, 2008A&A...482..333P}, only the (3:2 3:2 5:4) and (3:2 4:3 4:3) initial conditions are left as a viable options for the starting state of the solar system. 

As already discussed in section $2$, interactions between the cold outer disk and the outer-most ice-giant are largely independent of the starting condition, since scattering in a successful simulation always sets the planets onto orbits that are close to that of the current solar system, but with moderate eccentricities. Consequently, we did not restrict our analysis to any particular initial condition. Out of our set of $180$ integrations, in $8$ cases, primordially cold objects were able to retain unexcited orbits in addition to the gas-giant eigenmodes being reproduced correctly. Here, we focus on two representative integrations: one starting from the (2:1 4:3 4:3) initial condition (Figure 5) and another starting from the (5:3 4:3 3:2) initial condition (Figure 6). In both cases, the cold classical population is produced, but the wedge is only formed in the simulation that starts from the (2:1 4:3 4:3) initial condition (although it is somewhat smaller than its observed counterpart). Note, that the formation of the wedge has little to do with the initial condition - rather, its production is a random process. Similarly, the exact degree of excitation of the cold population's inclinations is sensitively dependent on the details of Neptune's evolution, which is chaotic. Thus, the fact that the wedge is reproduced in one simulation and the degree of excitation of the inclinations is reproduced in another are unrelated results. 

\begin{figure}[t]
\includegraphics[width=0.5\textwidth]{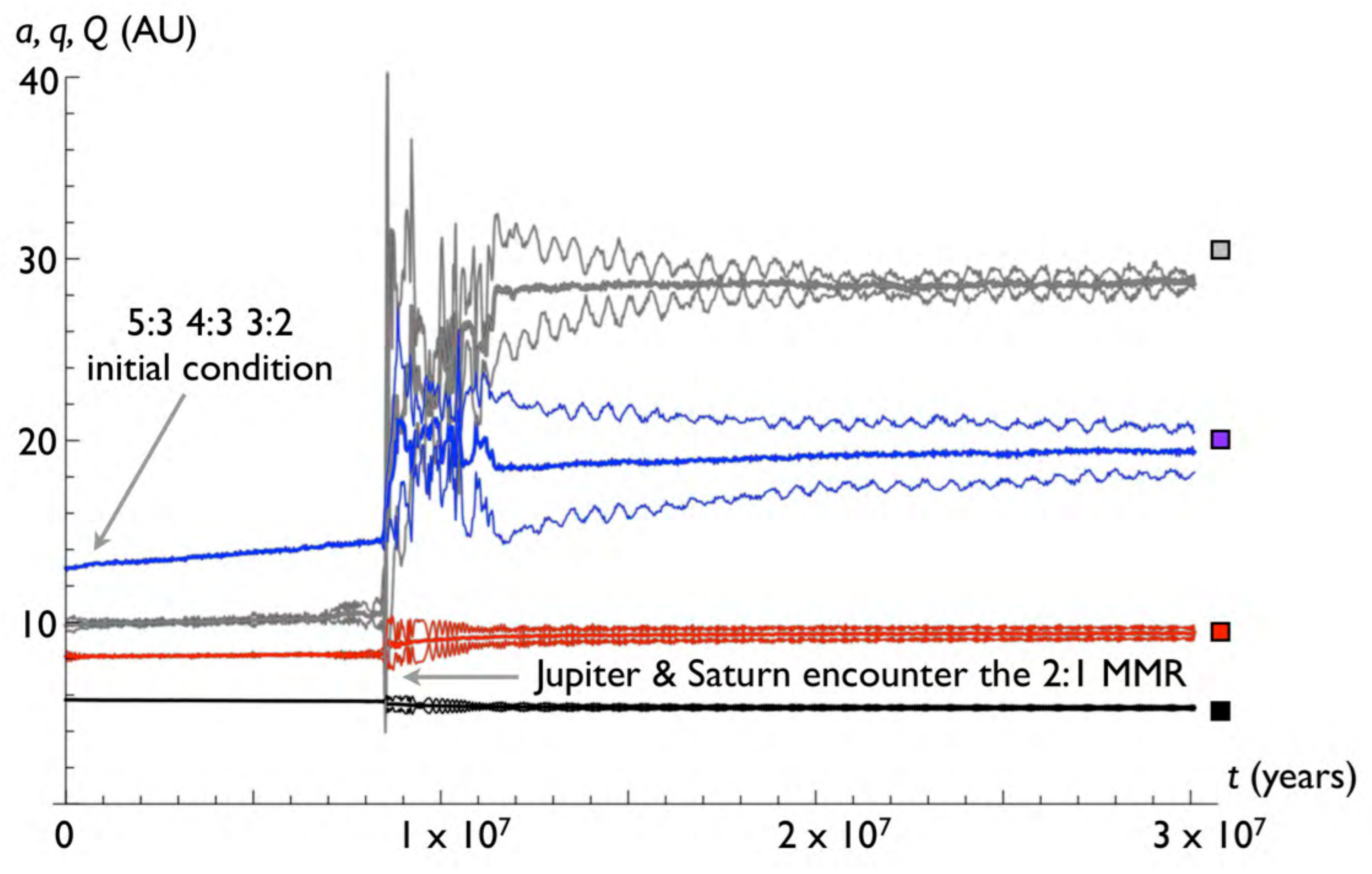}
\caption{Orbital evolution of planets. The system starts out in the (5:3 4:3 3:2) initial condition, and undergoes a brief period of instability when Saturn (Red) and Jupiter (Black) encounter a mutual 2:1 MMR. The boxes on the right of the plot correspond to actual semi-major axes of the giant planets. An evolved remnant planetesimal disk of this simulation is presented in Figures 9 \& 10.} 
\end{figure}

A vast majority ($>90\%$) of the objects in the cold classical region (i.e. $a \gtrsim 42$AU) are retained in our simulations on stable orbits. On the contrary, only about a few thousandth of the particles in the inner disk are emplaced onto stable orbits in the Kuiper belt region. This implies that in order to self-consistently study the formation of the Kuiper belt, $N \gg 3000$ is needed. Unfortunately, the required resolution is not computationally feasible. However, the problem can still be addressed by the use of ``tracer" simulations, an approach already utilized in the context of Kuiper belt formation by \cite{2008Icar..196..258L}. 

\begin{figure}[t]
\includegraphics[width=0.5\textwidth]{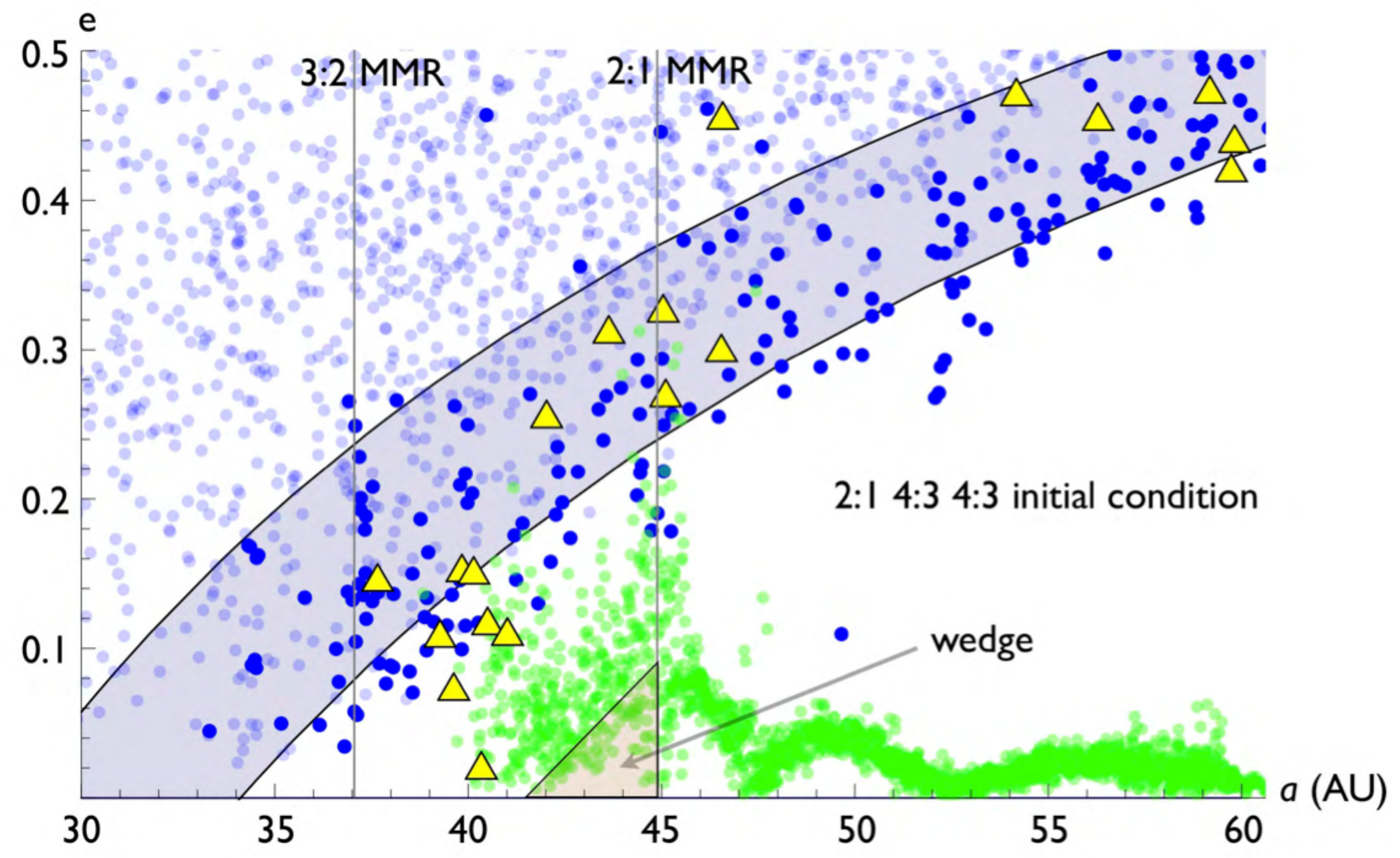}
\caption{Eccentricity distibution of the remnant planetesimal disk of the simulation that starts from the (2:1 4:3 4:3) multi-resonant state (see Figure 5). The pale blue dots show objects that originated interior to $\sim 35$AU, $30$Myr after the beginning of the simulation. The dark blue dots represent objects that originated interior to $35$AU, but are stable over $500$Myr. Green dots represent the test-particles that originate between $40$ \& $60$ AU. Yellow triangles represent test particles that originated between $35$ and $40$ AU. A wedge that is somewhat similar to the observed one (see Figure 1) forms in this simulation, as a result of a temporary slow-down in Neptune's precession rate (see Figure 11). Note that in this simulation, the classical Kuiper belt region lies between $\sim 37$ and $\sim 45$AU, as Neptune's final semi-major axis is $a \sim 28.3$AU. However, the aim here is to elucidate the physical mechanisms, rather than reproduce the actual Kuiper belt. The shaded region corresponds to the scattered disk.} 
\end{figure}

In a tracer simulation the planets and planetesimals are not self-consistently evolved in time. Rather, the evolution of the planets is pre-loaded from a master simulation and the planetesimals, which are treated as test particles, are evolved subject only to gravitational interactions with the planets. At the beginning of a tracer simulation, the tracer disk is initialized to have the same distribution as the massive component of the planetesimal disk. Consequently, at all times during the integration, the tracer particles also have an identical orbital distribution to that of the massive planetesimals. Each simulation was seeded with $100$ test particles and integrated on Caltech's \textit{PANGU} super-computer. We employed the Bulirsch-Stoer algorithm \citep{1992nrfa.book.....P} in our tracer integrations.

We performed 200 tracer simulations for each of the evolutions presented in Figures 5 \& 6. This amounts to evolving a primordial disk of $\sim 26000$ particles, including the outer belt. After the $\sim 30$ Myr simulations were completed, approximately $\sim 7\%$ of the particles that originated interior to $35$AU had semi-major axes in the range $35$AU $< a <$ $60$AU, shown as pale blue dots in Figures 7 - 10. We further cloned the populations\footnote{At the end of the simulations, there was only statistically significant structure in the $a,e.i$ distributions. The orbital angles took on random values during scattering.} of tracer particles in the Kuiper belt region to effectively increase the number of implanted hot classical, scattered, and resonant particles by another factor of $6$. The resulting Kuiper Belt, including the test-particles that originate beyond $35$AU, was then evolved for an additional 500 Myr to ensure that all unstable particles have time eject. At the end of the $500$Myr, only $\sim 5\%$ of the implanted objects, that were present at the end of the $30$Myr simulations, ended up on stable orbits. Consequently, the cumulative fraction of objects that are implanted into the Kuiper belt from the inner disk is $\sim 0.3 \%$. The stable objects are shown as dark blue dots in Figures 7 - 10.

\begin{figure}[t]
\includegraphics[width=0.5\textwidth]{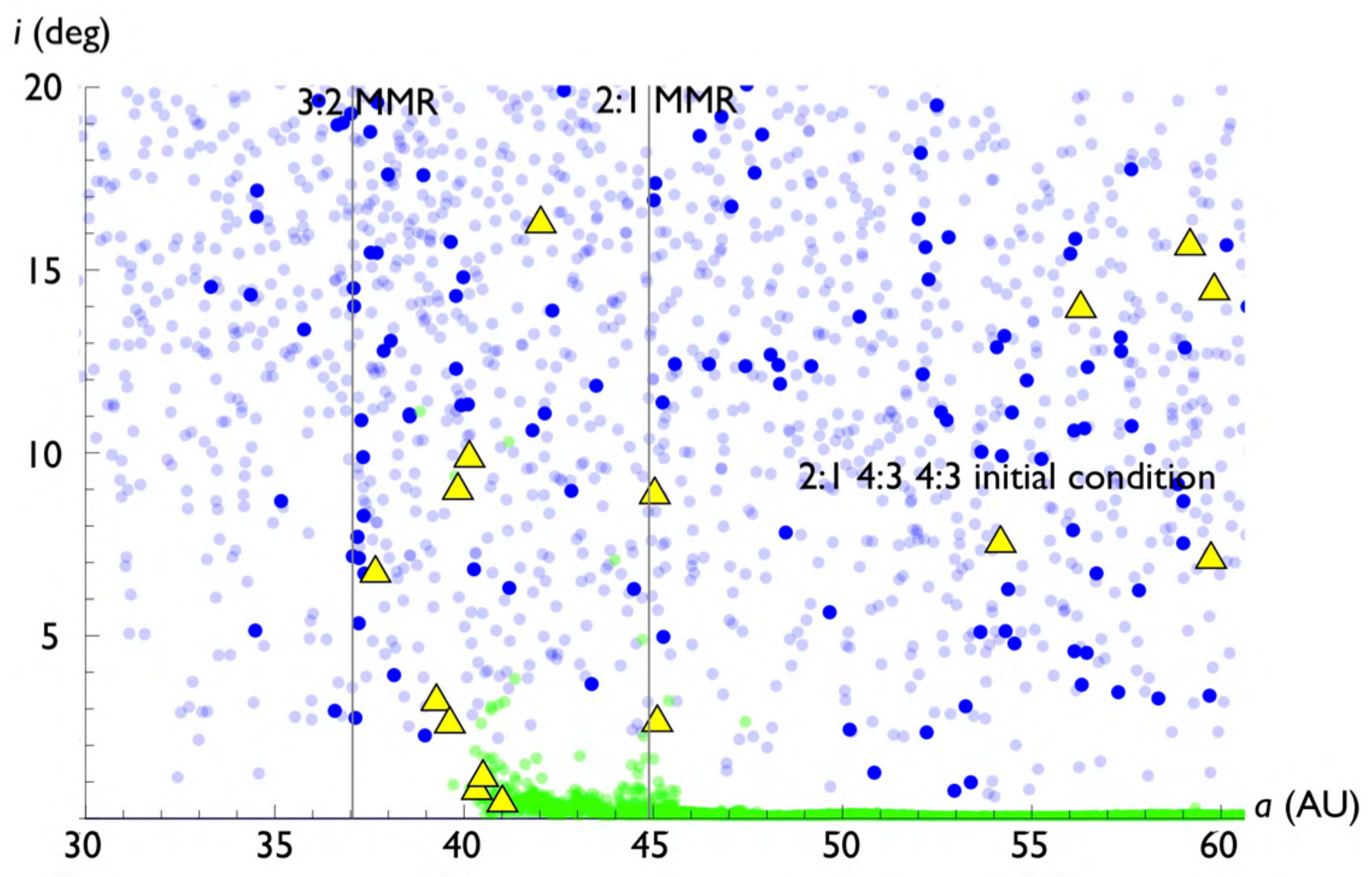}
\caption{Inclination distribution of the remnant planetesimal disk of the simulation that starts from the (2:1 4:3 4:3) multi-resonant state (see Figure 5). The pale blue dots show objects that originated interior to $\sim 35$AU, $30$Myr after the beginning of the simulation. The dark blue dots represent objects that originated interior to $35$AU, but are stable over $500$Myr. Green dots represent the test-particles that originate between $40$ \& $60$ AU. Yellow triangles represent test particles that originated between $35$ and $40$ AU.} 
\end{figure}

Note that in our simulations, the resonant populations are considerably diminished in number. This is largely a cost of performing self-consistent simulations with planetesimals that are unrealistically massive. Every time Neptune scatters a KBO, its resonances jump unrealistically far, disturbing the resonant KBOs, leading to their eventual ejection \citep{2006ApJ...651.1194M}. In a suite of customized simulations where the instability still occurs, but planets are analytically guided to their final orbits, and gravity is softened \citep{2008Icar..196..258L}, Neptune's MMR's end up overpopulated. This leads one to believe that the true parameter regime of Neptune's migration resided somewhere between what is presented in this work and that of \cite{2008Icar..196..258L} (Morbidelli, personal communication). 

Although both of the integrations presented here produce a cold classical belt, it is immediately apparent that the wedge is only produced in the integration that starts from the (2:1 4:3 4:3) initial condition, although again the process has little do with the choice of initial condition. Furthermore, from Figure 7, it can be readily inferred that the production of the wedge must be a secular effect since the structure in this simulation extends beyond the 2:1 MMR, i.e. the unswept region. Note that owing to the enhanced free precession of the KBOs (due to presence of a massive Kuiper belt), the wedge structure is shifted to the right, compared with analytical estimates presented in the previous section.

In the context of these integrations, we are further able to confirm that the formation of the wedge is due to a considerable slowdown in Neptune's precession. During circularization in the integration that starts from a (5:3 4:3 3:2) initial condition, Neptune precession is always roughly $g \approx 4.7 g_8$, while it is eccentric. On the contrary, in the integration that starts from the (2:1 4:3 4:3) initial condition, Neptune's precession rate varies considerably ($1.6 g_8 \lesssim g \lesssim 3.7g_8 $) between $23$Myr and $25$Myr (Figure 11). The presence of a mechanism for the successful formation of the wedge, from a local population is an important argument for confirmation of the in-situ formation of the cold classical population in the context of an instability model. 

\begin{figure}[t]
\includegraphics[width=0.5\textwidth]{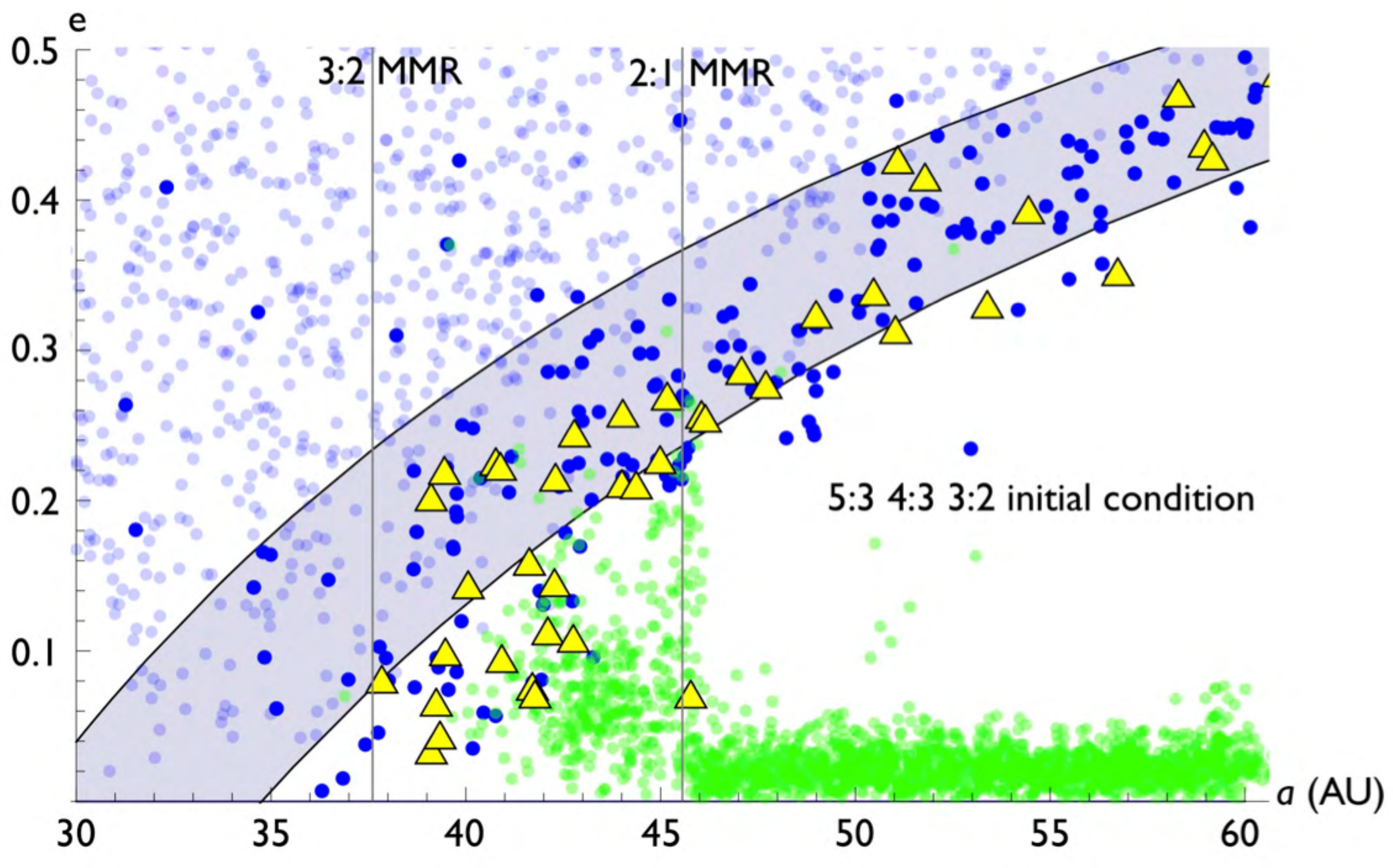}
\caption{Eccentricity distribution of the planetesimal disk of the simulation that starts from the (5:3 4:3 3:2) multi-resonant state (see Figure 5). The pale blue dots show objects that originated interior to $\sim 35$AU, $30$Myr after the beginning of the simulation. The dark blue dots represent objects that originated interior to $35$AU, but are stable over $500$Myr. Green dots represent the test-particles that originate between $40$ \& $60$ AU. Yellow triangles represent test particles that originated between $35$ and $40$ AU. Note that the wedge does not form in this simulation, because Neptune's precession never slows down, while it is eccentric. Note that in this simulation, the classical Kuiper belt region lies between $\sim 38$ and $\sim 46$AU, as Neptune's final semi-major axis is $a \sim 29$AU. However, the aim here is to elucidate the physical mechanisms, rather than reproduce the actual Kuiper belt. The shaded region corresponds to the scattered disk.} 
\end{figure}

It is noteworthy that in the results of the simulation, the wedge appears much less coherent, at semi-major axes interior to the 2:1 MMR. This is a consequence of eccentric resonant sweeping. Because of Neptune's considerable eccentricity, the KBO multiplet and the Neptune multiplet of the resonance overlap even for small KBO eccentricities. This allows the KBO to randomly explore the phase-space occupied by both sections of the resonance. However, as Neptune's eccentricity is monotonically decreasing, so is the phase space volume occupied by Neptune's multiplet of the 2:1 MMR, making capture impossible \citep{2006MNRAS.372L..14Q}. Moreover, because of different precession rates, the nominal location of Neptune's multiplet of the resonance lags (i.e smaller semi-major axis) that of the KBO. Thus, if a KBO exits the resonance shortly after it enters, it tends to get transported closer to the sun, since it enters at the KBO multiplet and exits at the Neptune multiplet. The change in semi-major axes, however is only the resonant splitting width, so it is rather small ( $\delta a < 0.1$AU). This randomization of the orbital elements causes the inner part of the wedge to appear less coherent in Figure 7. 

It is finally worth noting that although KBOs that become the cold classical population are able to roughly retain their primordial orbital distribution, the objects between $35$AU and $40$AU inevitably get scattered by Neptune during the instability. Indeed, in both simulations presented here, the scattered cold classicals (shown as yellow triangles in Figures 7-10) join the scattered disk as well as the hot classical population, while some particles get trapped in resonances temporarily, during their evolution. 

\begin{figure}[t]
\includegraphics[width=0.5\textwidth]{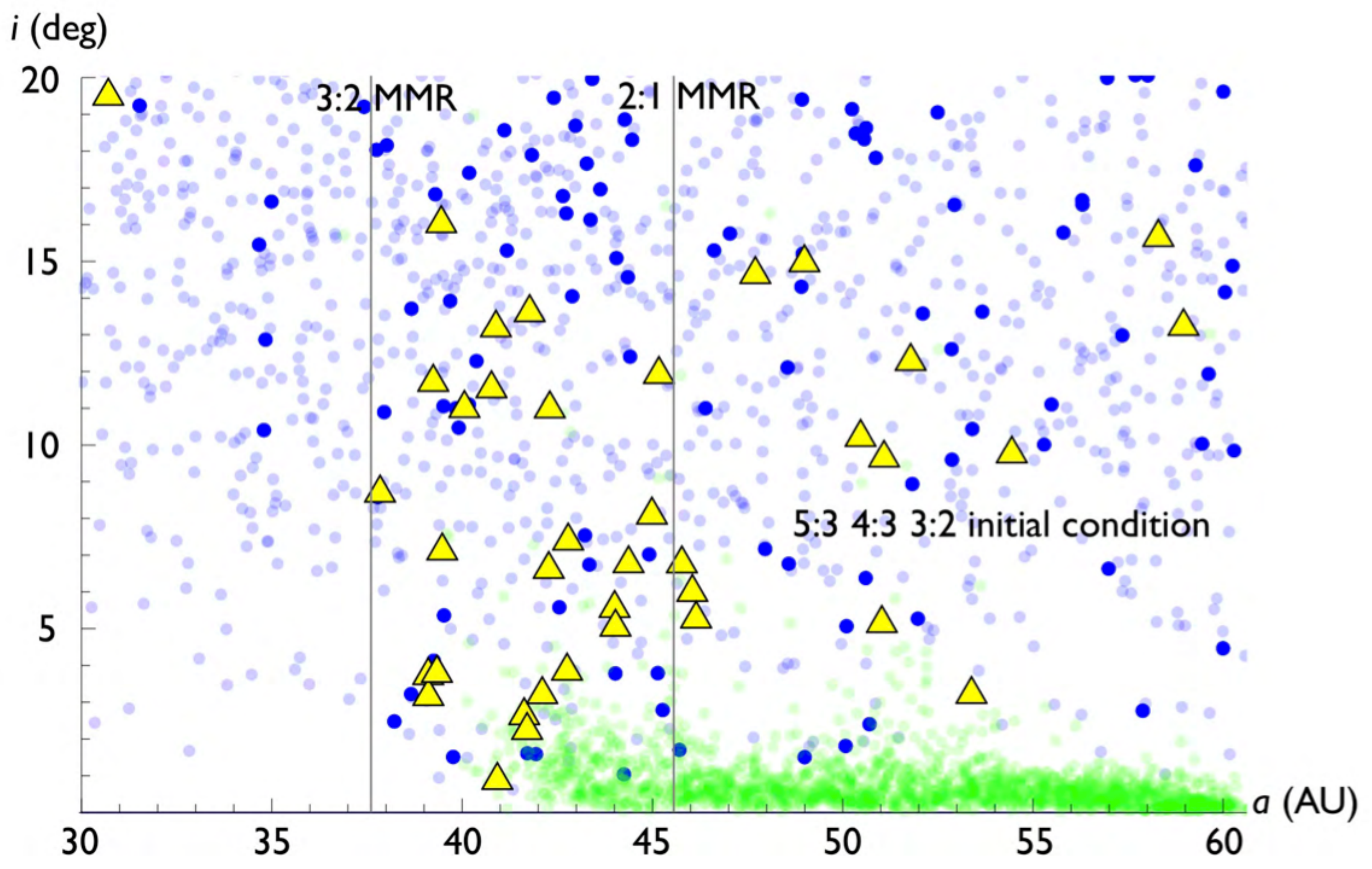}
\caption{Inclination distribution of the planetesimal disk of the simulation that starts from the (5:3 4:3 3:2) multi-resonant state (see Figure 5). The pale blue dots show objects that originated interior to $\sim 35$AU, $30$Myr after the beginning of the simulation. The dark blue dots represent objects that originated interior to $35$AU, but are stable over $500$Myr. Green dots represent the test-particles that originate between $40$ \& $60$ AU. Yellow triangles represent test particles that originated between $35$ and $40$ AU.} 
\end{figure}

The fact that these lifted objects mostly get emplaced onto stable orbits is suggestive that the results of intrusion of inclined populations by cold-classical like objects, that took place during the instability, should still be observable today. In other words, the in-situ formation scenario for cold-classicals presented here predicts that that a class of objects, occupying the same unique color region as the cold classicals, should be present in the excited populations. 

\section{Discussion}

In this paper, we present a self-consistent dynamical model for the evolution of a primordial cold classical population of the Kuiper belt, in the context of an instability-driven formation scenario for the solar system.  We show, from simple analytical considerations, that the cold belt can survive the transient period of dynamical instability, inherent to the planets. In order for a primordially cold population of KBOs to maintain an unexcited state, the average apsidal precession and nodal recession rates of Neptune during the transient phase of instability must have been considerably faster than what is observed in today's solar system. Simultaneously, successful formation of the wedge (see Figure 1) requires that the apsidal precession rate drops by a factor of a few for a short period of time. Numerical integrations presented in this work confirm the results of the analytical calculations and reveal a particular result, where the formed cold population and the wedge closely resemble their observed counter-parts. The dynamical evolution of cold classicals we propose here is in close agreement with the uniqueness of cold classical's physical characteristics. 

In situ formation of cold classicals brings to light the issue of truncation of the classical belt near the 2:1 MMR. In the chaotic capture mechanism proposed by \cite{2008Icar..196..258L}, the outer edge comes about naturally, as the 2:1 MMR sculpts the belt. In our solution, however, a cold belt that extends further out is surely possible. Thus, we are forced to attribute the proximity of the edge and the 2:1 MMR to a mere coincidence. Another question of interest is the fate of primordially cold binaries in the $35 - 40$AU region. It is likely that many of these binaries will get disrupted by close-encounters with Neptune, although the exact fraction will depend on the details of Neptune's evolution. Consequently, an in-depth analysis of the evolution of the scattered cold KBOs may open up an avenue towards further constraining the orbital history of Neptune. 

\begin{figure}[t]
\includegraphics[width=0.5\textwidth]{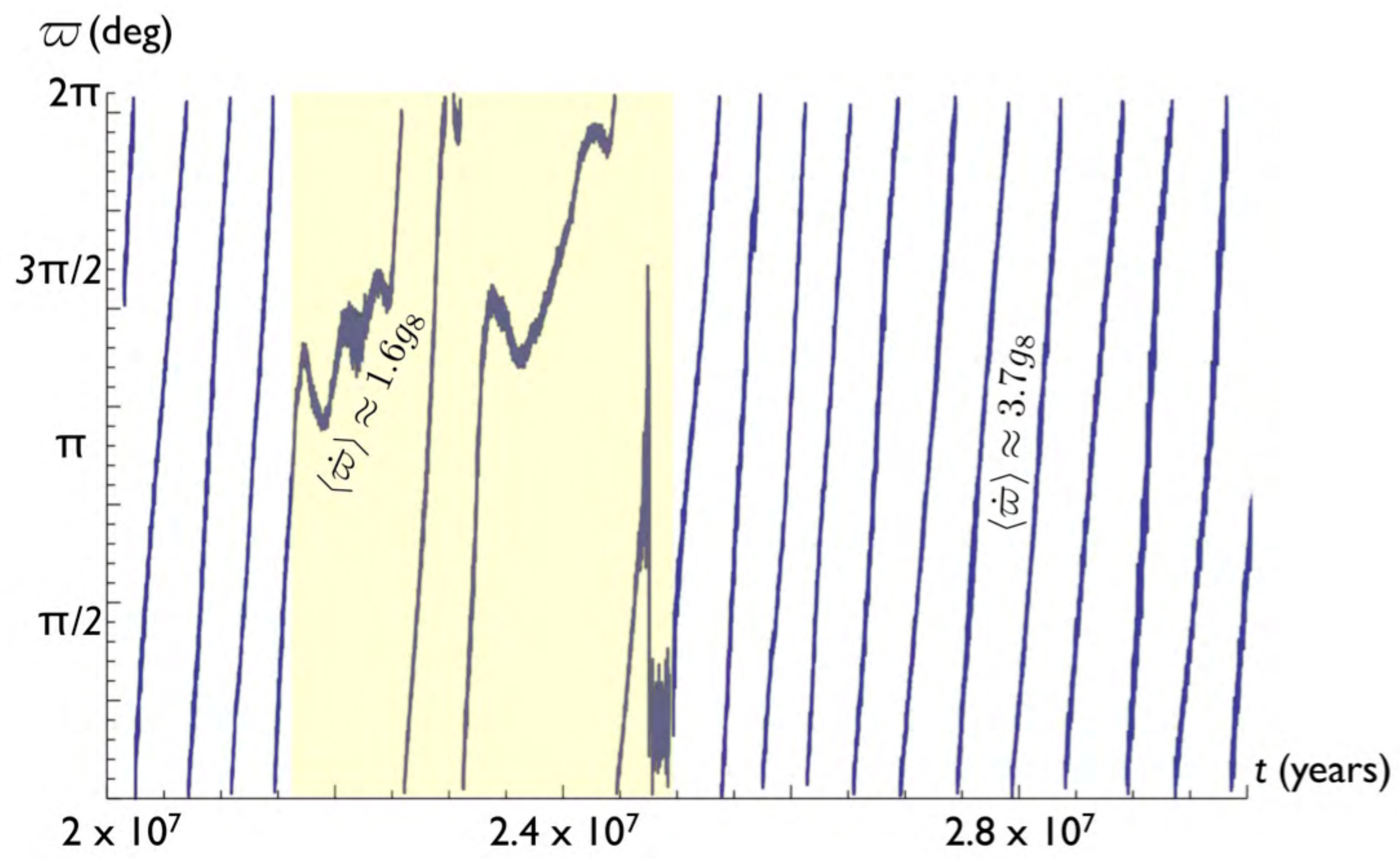}
\caption{Precession of Neptune's longitude of perihelion in the simulation that originates from the (2:1 4:3 4:3) multi-resonant initial condition (see Figure 5). Most of the time, Neptune's precession rate exceeds its current value by a factor of a few. However, the precession rate slows down considerably at $t \approx 22$Myr. The wedge forms as a result of the highlighted slowdown in Neptune's apsidal precession.} 
\end{figure}

Although in-situ formation of cold classicals resolves a pressing dynamical problem within the Nice model, it gives rise to a new issue that requires attention. Namely, the outstanding question of importance is planetesimal formation beyond $\sim 35$AU, given the steep size distribution of the cold classicals. In other words, how is the formation of planetesimals, up to $\sim 200$ km in size accomplished in such a low-density environment? 

Although the answer to this question is by no means trivial, one possible solution to this problem lies within the context of streaming instabilities \citep{2005ApJ...620..459Y}. Streaming instabilities have already been suggested as the dominant formation process in the cold classical population, as gravitational collapse has been shown to yield wide binaries \citep{2010AJ....140..785N}. Importantly, in the proposed picture, planetesimal formation is a threshold process, that ``turns on," only when gas drag accumulates a critical amount of dust in a given location within the solar nebula. Thus, one can in principle envision a system where most of the dust gets carried inward of $\sim 35$AU by gas drag, but infrequently, the dust surface density reaches a critical value in the outer nebula, causing a few, but sizable planetesimals to be born. Such a scenario would likely result in a very sharply decreasing surface density profile in the outer nebula, at the epoch of disappearance of the gas. As a result, this picture would imply the existence of a steep density gradient in the primordial planetesimal disk, such as the one we require in our model, consistent with preventing Neptune's extended migration.

Another possibility for the formation process is hierarchical coagulation, where planetesimal growth is accomplished by collisions among smaller objects in a quiescent environment \citep{2002PASP..114..265K}. Particularly, it has been suggested that if aided by turbulent concentration, hierarchical coagulation could yield the desired mass of the cold classical population \citep{2010Icar..208..518C}. In fact, even if the original mass of the cold belt exceeded its current value, erosion by collisional grinding could in principle be invoked to reduce the overall mass. However this process may prove problematic in reproducing the observed wide binary fraction of the cold belt  \citep{2011arXiv1102.5706N}.

Whatever the formation process for the cold classicals is, the results presented here have considerable implications. First and foremost, the successful retention of the cold-classical population in the context of an instability-driven model, fixes the most significant drawback of the Nice model. Second, our scenario suggests that the cold classical Kuiper belt is the only population of objects in the outer solar system that has not been transported away from its formation site. Furthermore, assuming that collisional grinding has played a negligible role in the cold population's evolution (as suggested by the observed binary fraction \citep{2011arXiv1102.5706N}), the cold classical population essentially yields the surface density of the solar nebula at $a \sim 45$AU, since the majority of the KBOs are retained in place. This potentially makes the cold classicals a unique laboratory for the study of surface processes as well as the chemistry of the primordial solar nebula. Third, based upon the results of the numerical simulations, we expect that objects that are physically similar to the cold classicals should be scattered throughout the Kuiper belt and as a result may explain the spectral similarity between cold classicals and some objects at higher inclinations. 

In conclusion, it appears that quantitative evaluation of planetesimal formation beyond $\sim 35$AU is required to draw a complete picture of the of the in-situ formation and evolution scenario for the cold classicals. However, the considerable improvement of the model for early dynamical evolution of the Kuiper belt presented here supports the overall validity of the hypothesis.\\

\textbf{Acknowledgments} We thank Alessandro Morbidelli, Hal Levison, Darin Ragozzine and Peter Goldreich for useful conversations.

\end{document}